\documentclass[10pt,letterpaper]{article}
\usepackage{opex3}
\usepackage{cite} % to render, for example, [5-8] instead of [5, 6, 7, 8].
\usepackage{amssymb} % to use the approximately-greater-than symbol

\begin{document}

%%%%%%%%%%%%%%%%%% title page information %%%%%%%%%%%%%%%%%%
\title{A versatile high resolution objective for imaging quantum gases}

\author{L.~M.~Bennie$^{*}$, P.~T.~Starkey, M.~Jasperse, C.~J.~Billington, R.~P.~Anderson and L.~D.~Turner}

\address{School of Physics, Monash University, Victoria 3800, Australia.}

\email{$^*$lisa.bennie@monash.edu} %% email address is required

%%%%%%%%%%%%%%%%%%% abstract and OCIS codes %%%%%%%%%%%%%%%%
%% [use \begin{abstract*}...\end{abstract*} if exempt from copyright]

\begin{abstract}
We present a high resolution objective lens made entirely from catalog singlets that has a numerical aperture of 0.36.
It corrects for aberrations introduced by a glass window and has a long working distance of 35\,mm, making it suitable for imaging objects within a vacuum system.
This offers simple high resolution imaging for many in the quantum gas community.
The objective achieves a resolution of 1.3\,$\textrm{\textmu m}$ at the design wavelength of 780\,nm, and a diffraction-limited field of view of 360\,$\textrm{\textmu m}$ when imaging through a 5\,mm window.
Images of a resolution target and a pinhole show quantitative agreement with the simulated lens performance.
The objective is suitable for diffraction-limited imaging on the D2 line of all the alkalis by changing only the aperture diameter, retaining numerical apertures above 0.32.
The design corrects for window thicknesses of up to 15\,mm if the singlet spacings are modified.
\end{abstract}


%%%%%%%%%%%%%%%%%%%%%%% References %%%%%%%%%%%%%%%%%%%%%%%%%

%%%%%%%%%%%%%%%%%%%%%%%%%%  body  %%%%%%%%%%%%%%%%%%%%%%%%%%

\section{Introduction}
Imaging systems are an essential component of quantum gas experiments.
There is an increasing need for high resolution imaging ($<3\,\textrm{\textmu m}$) to extract fine detail, such as topological defects in a Bose--Einstein condensate~\cite{kawaguchi_spinor_2012}, or a single trapped ion~\cite{streed_absorption_2012}.
One difficulty faced is that the object of interest is held within a vacuum chamber and must be viewed through a thick glass window.
Even a perfect optical flat introduces significant spherical aberration~\cite[p.\,572]{gross_handbook_2005} in a large numerical aperture (NA) system.

One approach is to house an infinity-corrected objective within the vacuum chamber~\cite{sortais_diffraction-limited_2007,bakr_quantum_2009}.
Collimated rays from such objectives are not significantly aberrated because they pass through an optically flat window at near-normal incidence.
With this approach the objective can have a short working distance, allowing for large NAs and permitting the use of commercial microscope objectives~\cite{salim_high-resolution_2012}.
However \textit{in vacuo} optics must be vacuum-compatible and bakeable, and are fixed in positions that limit optical access to the experiment.

Optics housed external to the vacuum chamber are particularly desirable, for example, in a Bose-Einstein condensate apparatus that shares a common source chamber and science chamber~\cite{lin_rapid_2009}, or in the study of liquid helium using optical cryostats~\cite{douillet_easy--build_2000}.
The \textit{ex vacuo} objective must have a long working distance if the object is located far ($\gtrsim 5\,\mathrm{mm}$) within a vacuum chamber and must also correct for the spherical aberrations introduced by the vacuum window.
Fortunately in quantum gas experiments the illumination is typically monochromatic with a small field of view ($\sim 1\,\mathrm{mm}$), nevertheless long working distance objectives that correct for large spherical aberrations are not available commercially and remain a design challenge.
Published designs that meet these criteria~\cite{alt_objective_2002, ottenstein_new_2006, nelson_imaging_2007, bucker_single-particle-sensitive_2009, sherson_single-atom-resolved_2010, zimmermann_high-resolution_2011} require manufacturing at least one custom singlet, which is a slow and costly process.
Because of this difficulty, many experiments employ a single lens, or a single light-gathering objective lens followed by a single image forming lens~\cite{anderson_spatial_1999}.
Such imaging systems are not corrected for aberrations and consequently have low resolution.

This article provides an alternative solution to the problem of high resolution imaging with \textit{ex vacuo} optics: an objective consisting of entirely catalog singlets that achieves diffraction-limited imaging with an NA of 0.36.
The objective achieves a resolution of 1.3\,$\textrm{\textmu m}$ with a field of view (FOV) of 360\,$\textrm{\textmu m}$ for a 5\,mm thick window using 780\,nm illumination.
The objective is suitable for diffraction-limited imaging of all the alkalis, retaining NAs above 0.32 by changing only the aperture diameter and accounting for the chromatic focal shift.
The objective remains well-corrected into the near ultraviolet; we predict sub-micrometer resolution when imaging Yb$^+$ in ion trapping experiments.
A useful diffraction-limited FOV can be retained for glass windows up to 15\,mm thick by changing the singlet spacings.
The required spacings can be found through re-optimization of the design in ray-tracing software.

\section{Design and construction}

\begin{table}[tb]
\caption{(a) Lens prescription of the objective suitable for entry into a ray-tracing system ($\lambda = 780\,\textrm{nm}$, 5\,mm silica window, 35\,mm working distance, $f = 47\,\textrm{mm}$).
(b) Changing only the aperture diameter enables diffraction-limited imaging of all the alkalis and Yb$^+$.}
\vspace{\baselineskip}
\centering
\small
\begin{tabular}[t]{l r@{.}l r@{.}l l}
\multicolumn{6}{l}{(a) Objective prescription}\\
\hline
Sur-  &\multicolumn{2}{l}{Curvature} &\multicolumn{2}{l}{Thick-}    &Mat-\\
face  &\multicolumn{2}{l}{rad. (mm)} &\multicolumn{2}{l}{ness (mm)} &erial\\
\hline
1     &\multicolumn{2}{c}{$\infty$}  &4&00                          &BK7\\
2     &51&50                         &10&50                         &air\\
3     &89&47                         &10&36                         &BK7\\
4     &-89&47                        &0&76                          &air\\
5     &47&90                         &7&30                          &BK7\\
6     &119&30                        &1&12                          &air\\
7     &30&30                         &9&70                          &BK7\\
8     &65&80                         &7&30                          &air\\
9     &\multicolumn{2}{c}{$\infty$}  &14&94                         &air\\
10    &\multicolumn{2}{c}{$\infty$}  &5&00                          &silica\\
11    &\multicolumn{2}{c}{$\infty$}  &15&00                         &vacuum\\
\hline
\end{tabular}
\ 
\begin{tabular}[t]{l l l l l l}  %r@{.}l
\multicolumn{6}{l}{(b) Objective performance} \\
\hline
           &$\lambda$ &Ap    &Res                    &NA    &FOV\\
           &(nm)      &(mm)  &($\textrm{\textmu m}$) &      &($\textrm{\textmu m}$)\\
\hline
Li         &671       & 24.0 & 1.20                  & 0.34 &400 \\
Na         &589       & 22.0 & 1.13                  & 0.32 &560 \\
K          &767       & 25.0 & 1.33                  & 0.35 &400 \\
Rb         &780       & 25.4 & 1.31                  & 0.36 &360 \\
Cs         &852       & 26.0 & 1.44                  & 0.36 &340 \\
Fr         &718       & 24.0 & 1.29                  & 0.34 &470 \\
Yb$^+$     &370       & 16.0 & 0.90                  & 0.25 &520 \\
\hline
\multicolumn{6}{l}{}\\
\multicolumn{6}{l}{Ap = aperture diameter.}\\
\multicolumn{6}{l}{Res = resolution}\\
\multicolumn{6}{l}{FOV = diffraction-limited FOV diameter}\\
\end{tabular}
\label{table}
\end{table}

Our Bose--Einstein condensate apparatus is similar to Ref.~\cite{lin_rapid_2009}, but with an optically flat glass cell as our science chamber to maximize optical access.
The magnetic coils surrounding the glass cell accommodate 50.8\,mm diameter lenses, while the distance between the 5\,mm thick glass window and the condensate demands a working distance greater than 25\,mm.
We require a diffraction-limited FOV of at least 300\,$\textrm{\textmu m}$ for imaging condensates \textit{in situ}.

The objective was designed using Zemax 12 SE~\cite{oslo_note}, a ray-tracing program which varies parameters to optimize the lens performance.
The design of Alt~\cite{alt_objective_2002} was chosen as a starting point because it achieved a diffraction-limited performance using BK7 glass---a material widely available in standard lens catalogs---with only four singlets.
Our design is distinguished from Alt's by a larger NA and the use of only catalog lenses.
After scaling the design to 50.8\,mm diameter singlets, we optimized both on- and 200\,$\textrm{\textmu m}$ off-axis for maximal NA with the required working distance.
The optimization routine chosen in Zemax minimized the squared sum of all wavefront aberrations up to 7th order.
The default merit function was modified to seek a diffraction-limited modulation transfer function (MTF).
This ensures high image contrast for objects with spatial frequencies up to the resolution determined by the NA~\cite[p.\,132]{gross_handbook_2007}.
Initially all singlet thicknesses, spacings and curvatures were varied until the diffraction limit was reached for the desired working distance and NA.
We then replaced the singlet most similar to a catalog lens, fixing its curvature and thickness to the catalog values, and repeated the optimization routine for the remaining lenses.
This process was repeated until all singlets were catalog lenses.
During this process we occasionally made dramatic changes to lens curvatures and spacings to ensure that the optimization routine did not stagnate in a local minimum~\cite{smith_modern_2005}.

\begin{figure}[tb]
\begin{center}
\includegraphics[width = 0.85\textwidth]{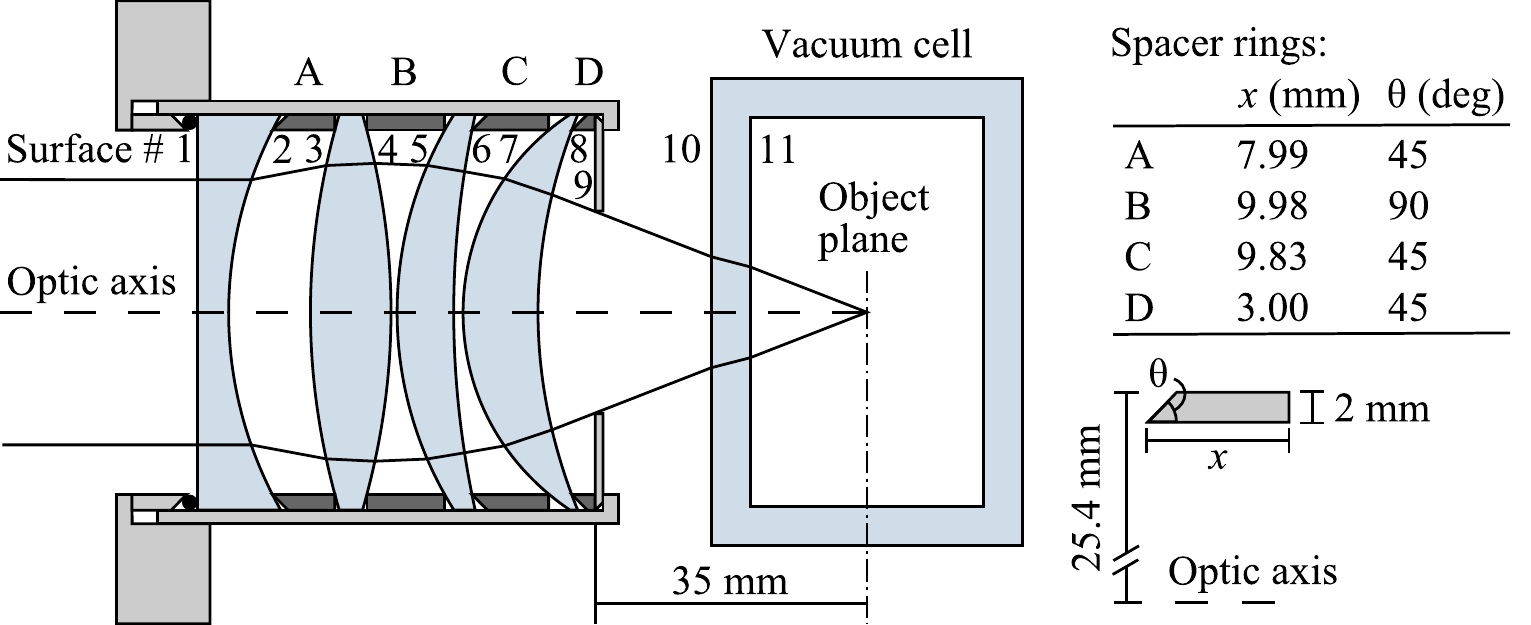}
\caption{Cross-section of the objective in its housing assembly.
The four catalog lenses, from left to right, are Thorlabs LC1093-B, Newport KBX151AR.16, Thorlabs LE1418-B and Thorlabs LE1076-B.
The cell is fused silica and is optically flat to less than $\lambda/4$.
The rays shown propagate from the object plane through the maximum aperture of the objective.
The geometry of the spacer rings A, B, C and D are shown.}
\label{lens_design}
\end{center}
\end{figure}

\begin{figure}[tb]
\begin{center}
\includegraphics[width = 0.5\textwidth]{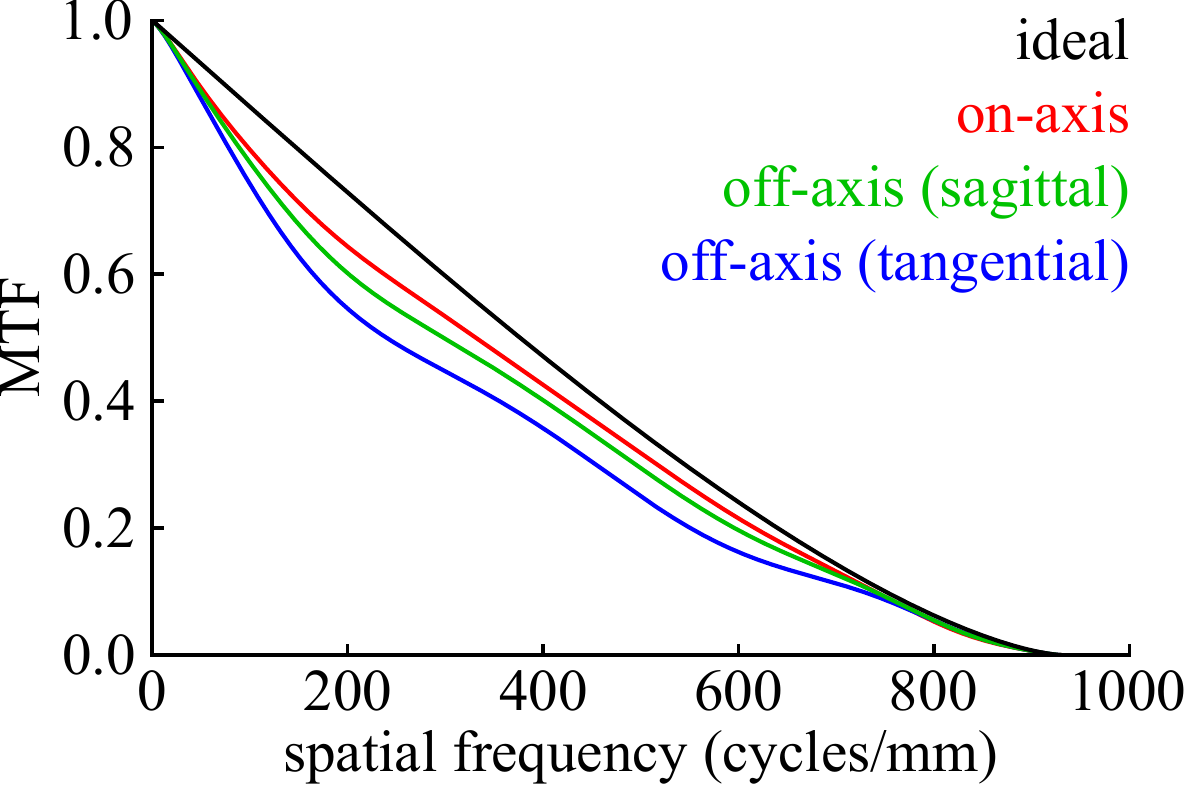}
\caption{The calculated MTF of the objective. On-axis and 180\,$\textrm{\textmu m}$ off-axis, both tangential and sagittal, exhibit marginally lower contrast than an aberration-free lens with the same NA.
Residual aberrations are a consequence of limiting the design to catalog singlets.}
\label{mtf}
\end{center}
\end{figure}

The final design comprises four catalog singlets: two positive meniscus lenses to enable a high NA with minimal addition of spherical aberration, a bi-convex lens, and a rear plano-concave lens to cancel aberrations introduced by the other lenses and by the window~\cite{fischer_spherical_1987}.
This design is shown in Figure~\ref{lens_design}, with the singlet curvatures and spacings listed in Table~\ref{table}(a).
Our objective has an NA of 0.36 yielding a resolution of 1.3\,$\textrm{\textmu m}$ at the design wavelength of 780\,nm.
It has a long working distance of 35\,mm measured from the front aperture (surface 9, Fig.~\ref{lens_design}) and an effective focal length of 47\,mm.
The objective is infinity-corrected, allowing the image-forming optics to be chosen separately to suit various applications.
The aberration produced by the 5\,mm thick window is corrected over a diffraction-limited FOV of 360\,$\textrm{\textmu m}$.
Here we consider the FOV to be `diffraction-limited' where the Strehl ratio is greater than~0.8~\cite[p.\,90]{gross_handbook_2007}.
The calculated MTF, both on-axis and at the edge of the FOV, is compared to the diffraction-limited MTF in Figure~\ref{mtf}.
The similarities between the curves indicate that the residual aberrations will not significantly degrade image contrast across the 360\,$\textrm{\textmu m}$ diameter FOV.

Our objective is applicable to imaging all the alkalis and Yb$^+$ at commonly used wavelengths.
Retaining a diffraction-limited FOV of a few hundred micrometers only requires changing the aperture diameter (Table (\ref{table}b)).
The objective is not corrected for chromatic aberration, so the focal shift must be accounted for when using a different wavelength.

While a full tolerance analysis was not performed, the design appears robust to changes in singlet spacing: rounding the spacings (thickness of surfaces 2, 4 and 6 in Table~\ref{table}) to the nearest 0.5\,mm had negligible effect on the predicted performance.
This allowed the housing assembly to be the simple design shown in Figure~\ref{lens_design}.
The singlets are held on-axis inside a smooth aluminium tube and are separated by aluminium spacer rings.
These rings contact each singlet 2\,mm from its outer edge, avoiding chamfers which can result in large spacing errors.
The tube screws into a mounting bracket to hold the singlets in place and an o-ring moderates the pressure applied to the singlets.
Aluminium components are anodized black to reduce reflections.

\section{Experimental performance}
We measured the MTF, FOV and point spread function of the objective using a USAF 1951 resolution target (Edmund Optics 58-198) and a 1\,$\textrm{\textmu m}$ pinhole (Edmund Optics 39-878), both illuminated by collimated 780\,nm laser light.
A 5\,mm fused silica optical flat was used to mimic a vacuum window.
An ${f=1000}$\,mm achromat placed immediately after the objective formed images of the test objects at a magnification of -21.4 on an sCMOS camera (Andor Neo).
This camera has a pixel size of 6.50\,$\textrm{\textmu m}$, corresponding to an effective pixel size of 304\,nm in the object plane.
Including the achromat in our simulation further compensated aberrations, permitting an increase of both the FOV to 400\,$\textrm{\textmu m}$ and the aperture diameter to 26\,mm.
Accordingly, this aperture diameter was used in the constructed objective.

\begin{figure}[tb]
\begin{center}
\includegraphics[width = 0.6\textwidth]{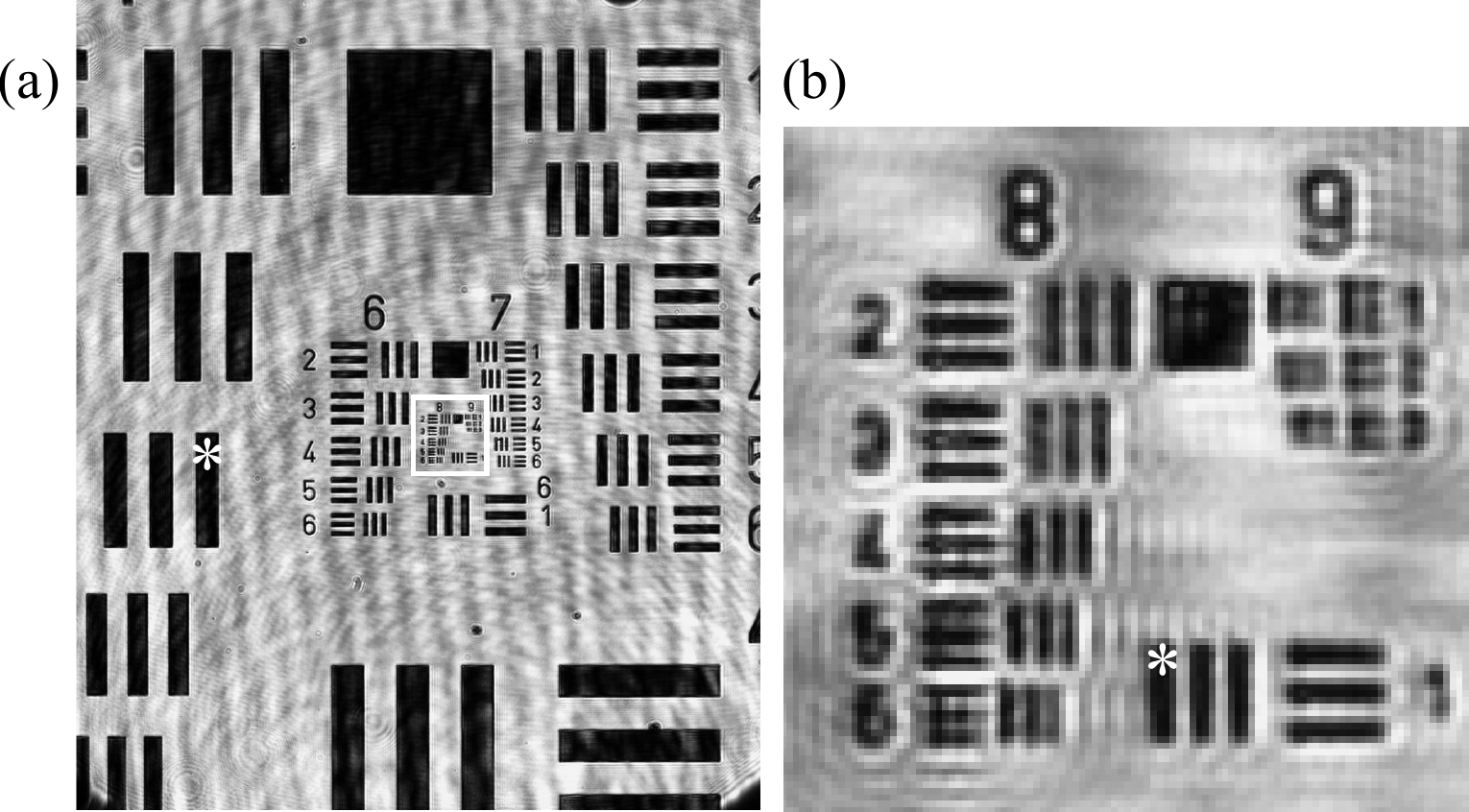}
\caption{Image of a USAF 1951 resolution target formed by the objective and an ${f=1000}$\,mm achromat.
The white rectangle in (a) marks the region shown in (b).
The bars marked with asterisks are (a) 22.1\,$\textrm{\textmu m}$ and (b) 1.95\,$\textrm{\textmu m}$ wide in the object plane.}
\label{target}
\end{center}
\end{figure}

The image of the resolution target (Fig.~\ref{target}) is undistorted beyond the diffraction-limited FOV.
The line pairs in element 6 of group 8 are clearly resolved, corresponding to a resolution of $\le 2.20\,\textrm{\textmu m}$.
The predicted resolution of 1.3\,$\textrm{\textmu m}$ would also resolve the elements in group 9, but the coherent illumination produced diffraction fringes which degraded the image quality.

\begin{figure}[tb]
\begin{center}
\includegraphics[width = 0.80\textwidth]{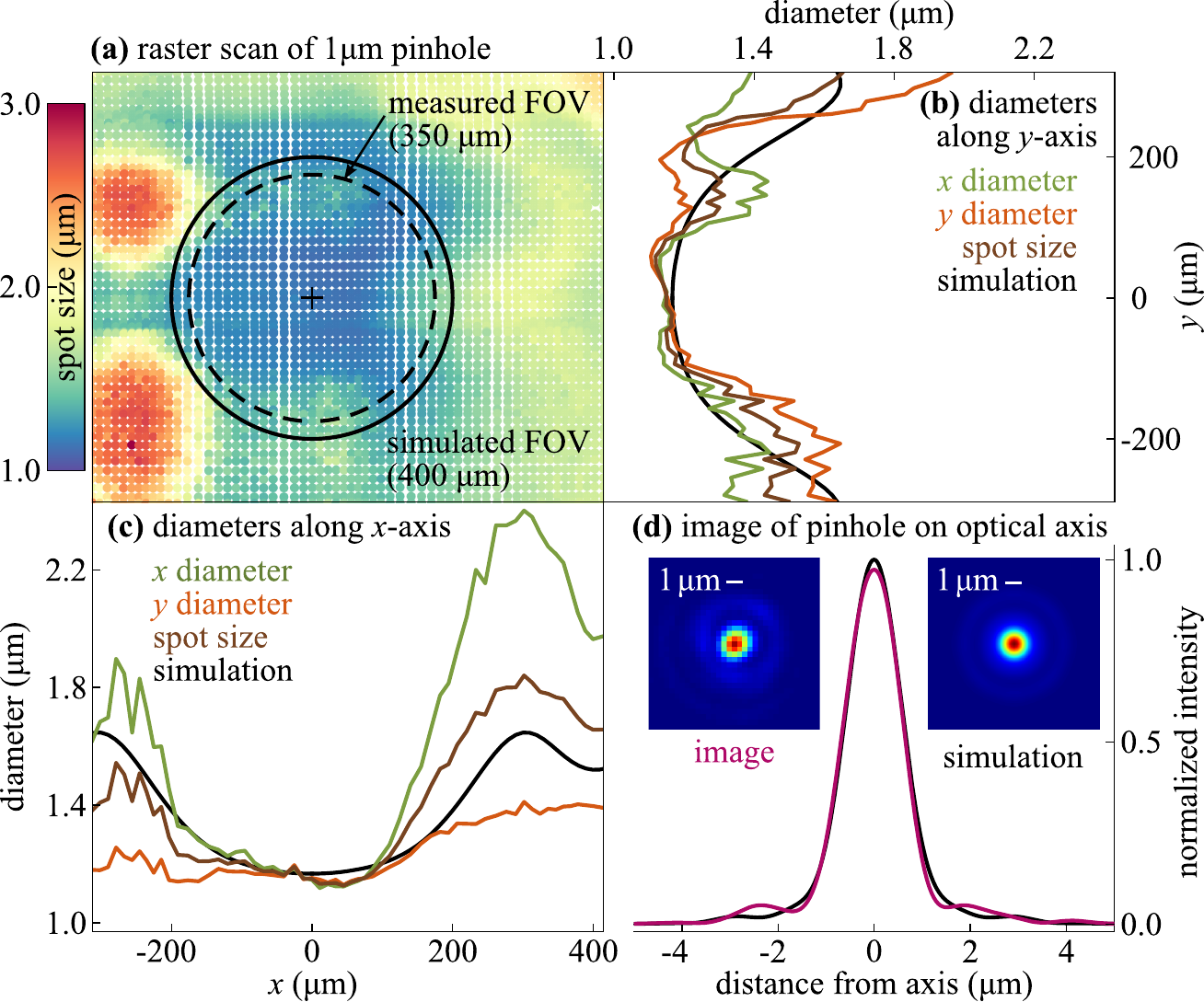}
\caption{(a) The spot size of a 1\,$\textrm{\textmu m}$ pinhole translated across the object plane to the edges of the camera chip.
(b, c) The spot size and constituent diameters across the $y$- and $x$-axis respectively through the optic axis (cross in (a)).
The measured spot size is in good agreement with our simulation within the FOV.
(d) The on-axis pinhole image compared to the simulated point spread function convolved with a 1\,$\textrm{\textmu m}$ pinhole.
The curves are the azimuthal averages of the inset images.}
\label{pinhole}
\end{center}
\end{figure}

The point spread function---the image of a point source---provides another measure of the lens performance.
The measured image of a 1\,$\textrm{\textmu m}$ pinhole is similar to the convolution of a~1\,$\textrm{\textmu m}$ top-hat function with the simulated point spread function (Fig.~\ref{pinhole}(d), inset).
The azimuthal averages of these images about the pinhole center (Fig.~\ref{pinhole}) are in agreement, affirming the objective performed as expected.
Asymmetry in the on-axis pinhole image is likely due to tilt of the optical flat or objective relative to the camera~\cite{ottenstein_new_2006}.

The pinhole was also used to measure the diffraction-limited FOV and to locate the position of the optic axis relative to the camera chip.
We raster-scanned the pinhole across the object plane using two motorized translation stages under automated control~\cite{starkey_scripted_2013}, taking 3600 images on a 10\,$\textrm{\textmu m}$ grid in one hour.
At each point on the grid we fitted a two dimensional Gaussian to the pinhole image; the geometric mean of the fitted rms diameters (spot size) are shown in Figure~\ref{pinhole}(a).
Row- and column-wise averages of the spot size revealed distinct minima, which we took to be the position of the optic axis (cross in Fig.~\ref{pinhole}(a)).
The diffraction-limited FOV was measured to be 350\,$\textrm{\textmu m}$ in diameter, commensurate with a circle inside which the spot size is less than $\sqrt{2}$ times the on-axis spot size.
The measured FOV is smaller than the simulated 400\,$\textrm{\textmu m}$, owing to the asymmetry in spot size about the optic axis.
This asymmetry was likely caused by tilt in the translation stages that moved the pinhole through either side of focus across the object plane.
Figures~\ref{pinhole}(b) and \ref{pinhole}(c) show the diameters of the pinhole image along the column and row that intersect the optic axis respectively.
In addition to the spot size, the constituent $x$ and $y$ Gaussian rms diameters are shown.
The variation between the $x$ and $y$ diameters reveals small astigmatism, suggesting tilt of the optical flat and objective relative to the camera.
The measured spot size compares favorably to simulation both on- and off-axis.

\section{Conclusion}
The combination of catalog singlets presented in this paper form a high resolution, long working distance objective suitable for imaging objects far within a vacuum chamber through a thick window.
Optical tests confirm the performance predicted using ray-tracing software.
Remarkably this objective enables diffraction-limited imaging across a wide FOV of all the alkalis and ytterbium ions, by changing only the aperture diameter.
Spherical aberration from a range of window thicknesses can be corrected by changing the singlet spacings.
The versatility and simplicity of this design makes it applicable to many experiments in the field of quantum gases.

\section{Acknowledgments}
This work was supported by Australian Research Council grants DP1094399 and DP1096830, and by a Monash University Faculty of Science equipment grant.


\begin{thebibliography}{10}
\bibitem{oslo_note}
The objective can also be modeled using the free software OSLO EDU, which is limited to ten surfaces, because the first surface is flat and need not be included in the model.
\end{thebibliography}
\end{document}